\documentclass[11pt]{article}
\usepackage{psfig} 
\usepackage{epsfig} 
\usepackage{epsf}

\def\ifmath#1{\relax\ifmmode #1\else $#1$\fi}

\def\3quarter{{\textstyle{3 \over 4}}}

\def\lf{\leaders\hbox to 1em{\hss.\hss}\hfill}
\def\e6{$E(6)$} 
\def\10{$SO(10)$}
\def\21{$SU(2) \otimes U(1) $}

\def\422{$SU(4) \otimes SU(2) \otimes SU(2)$} 
\def\321{$SU(3) \otimes SU(2) \otimes U(1)$}
\def\ne{\hbox{$\nu_e$ }}
\def\nm{\hbox{$\nu_\mu$ }}
\def\nt{\hbox{$\nu_\tau$ }}
\def\ns{\hbox{$\nu_{s}$ }}

\def\mnt{\hbox{$m_{\nu_\tau}$ }}

\def\neus{\hbox{neutrinos }}

\def\neu{\hbox{neutrino }}

\def\fig#1{{Fig. (\ref{#1})}}

\def\lsim{\raise0.3ex\hbox{$\;<$\kern-0.75em\raise-1.1ex\hbox{$\sim\;$}}}
\def\gsim{\raise0.3ex\hbox{$\;>$\kern-0.75em\raise-1.1ex\hbox{$\sim\;$}}}

\def\beq{\begin{equation}}
\def\eeq{\end{equation}}
\def\bef{\begin{figure}}
\def\eef{\end{figure}}
\def\bet{\begin{table}}
\def\eet{\end{table}}
\def\bea{\begin{eqnarray}}
\def\ba{\begin{array}}
\def\ea{\end{array}}
\def\bi{\begin{itemize}}
\def\ei{\end{itemize}}
\def\ben{\begin{enumerate}}
\def\een{\end{enumerate}}

\def\eea{\end{eqnarray}}
\def\apj#1#2#3{          {\it Astrophys. J. }{\bf #1} (19#2) #3}

\def\nat#1#2#3{          {\it Nature }{\bf #1} (19#2) #3}
\def\nps#1#2#3{        {\it Nucl. Phys. B (Proc. Suppl.) }{\bf #1} (19#2) #3} 
\def\np#1#2#3{           {\it Nucl. Phys. }{\bf #1} (19#2) #3}
\def\pl#1#2#3{           {\it Phys. Lett. }{\bf #1} (19#2) #3}
\def\pr#1#2#3{           {\it Phys. Rev. }{\bf #1} (19#2) #3}

\def\prl#1#2#3{          {\it Phys. Rev. Lett. }{\bf #1} (19#2) #3}

\def\zp#1#2#3{           {\it Zeit. fur Physik }{\bf #1} (19#2) #3}

\def\n.c.#1#2#3{         {\it Nuovo Cim. }{\bf #1} (19#2) #3}
\def\r.n.c.#1#2#3{       {\it Riv. del Nuovo Cim. }{\bf #1} (19#2) #3}

\def\mpl#1#2#3{          {\it Mod. Phys. Lett. }{\bf #1} (19#2) #3}

\def\ppnp#1#2#3{           {\it Prog. Part. Nucl. Phys. }{\bf #1} (19#2) #3}

\def\ip{in preparation}
\def\bne{\hbox{$\bar\nu_e$ }}  
\def\bnm{\hbox{$\bar\nu_\mu$ }}  
 
\begin{document}
\begin{center}

\textbf{\large Neutrino Physics at the Turn of the Millenium}

\emph{ Jos\'e W. F. Valle \\}
\textit{Instituto de F\'{\i}sica Corpuscular -- C.S.I.C.\\
  Departamento de F\'{\i}sica Te\`orica -- Univ. de
  Val\`encia, \\
  Edificio Institutos de Paterna, Apartado de Correos 2085,
  46071, Val\`encia \\
  http://neutrinos.uv.es }
\end{center}


\vspace*{0.9cm} 
\begin{abstract}
  
  I discuss the implications of the latest data on solar and
  atmospheric neutrinos which strongly indicate the need for physics
  beyond the Standard Model.  I review the theoretical options for
  reconciling these data in terms of three-neutrino oscillations.
  Even though not implied by the data, bi-maximal models of neutrino
  mixing emerge as an attractive possibility.  Supersymmetry with
  broken R--parity provides a predictive way to incorporate it,
  opening the possibility of testing neutrino anomalies at high--energy
  collider experiments such as the LHC or at the upcoming
  long-baseline or neutrino factory experiments.  Reconciling, in
  addition, the hint provided by the LSND experiment requires a
  fourth, light sterile neutrino.  The simplest theoretical scenarios
  are the most symmetric ones, in which two of the four neutrinos are
  maximally-mixed and lie at the LSND scale, while the others are at
  the solar mass scale.  The lightness of the sterile neutrino, the
  nearly maximal atmospheric neutrino mixing, and the generation of
  $\Delta {m_\odot^2}$ \& $\Delta {m_{atm}^2}$ all follow naturally
  from the assumed lepton-number symmetry and its breaking.  These two
  basic schemes can be distinguished at neutral-current-sensitive
  solar \& atmospheric neutrino experiments such as the Sudbury
  Neutrino Observatory.  However, underground experiments have
  \emph{not yet proven} neutrino masses, since there is a variety of
  alternative mechanisms. For example, flavour changing interactions
  can play an important r\^ole in the explanation of solar and of
  contained atmospheric data and could be tested through effects such
  as $\mu \to e + \gamma$, $\mu-e$ conversion in nuclei, unaccompanied
  by neutrino-less double beta decay.  Conversely, the room is still
  open for heavy unstable neutrinos. A short-lived \nm might play a
  r\^ole in the explanation of the atmospheric data.  Finally, in the
  presence of a sterile neutrino \ns, a long-lived \nt in the MeV range
  could delay the time at which the matter and radiation contributions
  to the energy density of the Universe become equal, reducing the
  density fluctuations on the smaller scales, and rescuing the
  standard cold dark matter scenario for structure formation.  In this
  case the light \ne, \nm and \ns would account for the solar \&
  atmospheric data.

\end{abstract}

\section{Introduction}
\vskip .1cm

Undoubtedly the solar~\cite{Cl99,Ga99,sksol504,sksol825} and
atmospheric ~\cite{atmexpold,atm52kt} neutrino problems provide the
two most important milestones in the search for physics beyond the
Standard Model (SM). Of particular importance has been the recent
confirmation by the Super-Kamiokande collaboration \cite{atm52kt} of
the zenith-angle-dependent deficit of atmospheric neutrinos.
Altogether solar and atmospheric data give a strong evidence for \ne
and \nm conversions, respectively.
Neutrino conversions are a natural consequence of theories beyond the
Standard Model~\cite{fae}.  The first example is oscillations of
small-mass neutrinos. The simplest way to account for the lightness of
neutrinos is in the context of Majorana neutrinos: their mass violates
lepton number.  Its most obvious consequences would be processes such
as neutrino-less double beta decay~\cite{Schechter:1982bd}, or CP
violation properties of neutrinos~\cite{Schechter:1981gk}, so far
unobserved. Neutrino masses could be hierarchical, with the light \nt
much heavier than the \nm and \ne. While solar neutrino rates favour
the small mixing angle (SMA) MSW solution, present data on the
recoil-electron spectrum prefer the large mixing MSW~\cite{MSW} (LMA)
solution~\cite{valencia:1999}.
When interpreted in terms of neutrino oscillations, the observed
atmospheric neutrino zenith-angle-dependent deficit clearly indicates
that the mixing involved is maximal. In short we have the intriguing
possibility that, unlike the case of quarks, neutrino mixing is
bi-maximal. 
Supersymmetry with broken R--parity provides an attractive origin for
bi-maximal neutrino oscillations, which can be tested not only at the
upcoming long-baseline or neutrino factory experiments but also at
high--energy collider experiments such as the LHC.

One should however bear in mind that there is a variety of alternative
solutions to the neutrino anomalies. Just as an example let me stress
the case for lepton flavour violating neutrino transitions, which can
arise without neutrino masses
~\cite{Valle:1987gv,Guzzo:1991cp,Nunokawa:1996tg}.  They may still fit
present solar ~\cite{Krastev:1997cp} and contained atmospheric
~\cite{Gonzalez-Garcia:1998hj} data pretty well. They may arise in
models with extra heavy leptons \cite{WYLER,SST,BER,3E} and in
supergravity theories ~\cite{Hall:1986dx}.  A possible signature of
theories leading to FC interactions would be the existence of sizeable
flavour non-conservation effects, such as $\mu \to e + \gamma$,
$\mu-e$ conversion in nuclei, unaccompanied by neutrino-less double
beta decay if neutrinos are massless. In contrast to the intimate
relationship between the latter and the non-zero Majorana mass of
neutrinos due to the Black-Box theorem~\cite{Schechter:1982bd} there
is no fundamental link between lepton flavour violation and neutrino
mass. Other possibilities involve neutrino decays~\cite{V} and
transition magnetic moments~\cite{SFP} coupled to either to
regular~\cite{RSFP,akhmedov97} or to random magnetic
fields~\cite{Semikoz:1998xu}.

In addition to the solar and atmospheric neutrino data from
underground experiments there is also some indication for neutrino
oscillations from the LSND experiment~\cite{LSND,Louis:1998qf}. 
Barring exotic neutrino conversion mechanisms one requires {\sl three
  mass scales} in order to reconcile all of these hints, hence the
need for a light sterile neutrino~\cite{ptv92,pv93,cm93}. Out of the
four neutrinos, two of them lie at the solar neutrino scale and the
other two maximally-mixed neutrinos are at the HDM/LSND scale. The
prototype models proposed in~\cite{ptv92,pv93} enlarge the \21 Higgs
sector in such a way that neutrinos acquire mass radiatively, without
unification nor seesaw.  The LSND scale arises at one-loop, while the
solar and atmospheric scales come in at the two-loop level, thus
accounting for the hierarchy. The lightness of the sterile neutrino,
the nearly maximal atmospheric neutrino mixing, and the generation of
the solar and atmospheric neutrino scales all result naturally from
the assumed lepton-number symmetry and its breaking.  Either \ne- \nt
conversions explain the solar data with \nm- \ns oscillations
accounting for the atmospheric deficit~\cite{ptv92}, or else the
r\^oles of \nt and \ns are reversed ~\cite{pv93}. These two basic
schemes have distinct implications at future solar \& atmospheric
neutrino experiments with good sensitivity to neutral current neutrino
interactions. Cosmology can also place restrictions on these
four-neutrino schemes~\cite{Raffelt:1999zg}.

\section{Indications for New Physics}
\vskip .1cm

The most solid hints in favour of new physics in the neutrino sector
come from underground experiments on
solar~\cite{Cl99,Ga99,sksol504,sksol825} and atmospheric
~\cite{atmexpold,atm52kt} neutrinos.  The most recent data correspond
to 825--day solar~\cite{sksol825} and 52 kton-yr atmospheric data samples,
respectively~\cite{atm52kt}.

\subsection{Solar Neutrinos}
\vskip .1cm
 
The solar neutrino event rates recorded at the radiochemical
Homestake, Gallex and Sage experiments are summarized as: $2.56 \pm
0.22$ SNU (chlorine), $72.3 \pm 5.6$ SNU (Gallex and
Sage)~\cite{Cl99,Ga99}. Note that only the gallium experiments are
sensitive to the solar $pp$ neutrinos.  On the other hand the $^8$B
flux from Super-Kamiokande water Cerenkov experiment is $(2.44 \pm
0.08) \times 10^6 {\rm cm^{-2} s^{-1} }$ ~\cite{sksol825}.
\begin{figure}[t]
\centerline{\protect\hbox{
\psfig{file=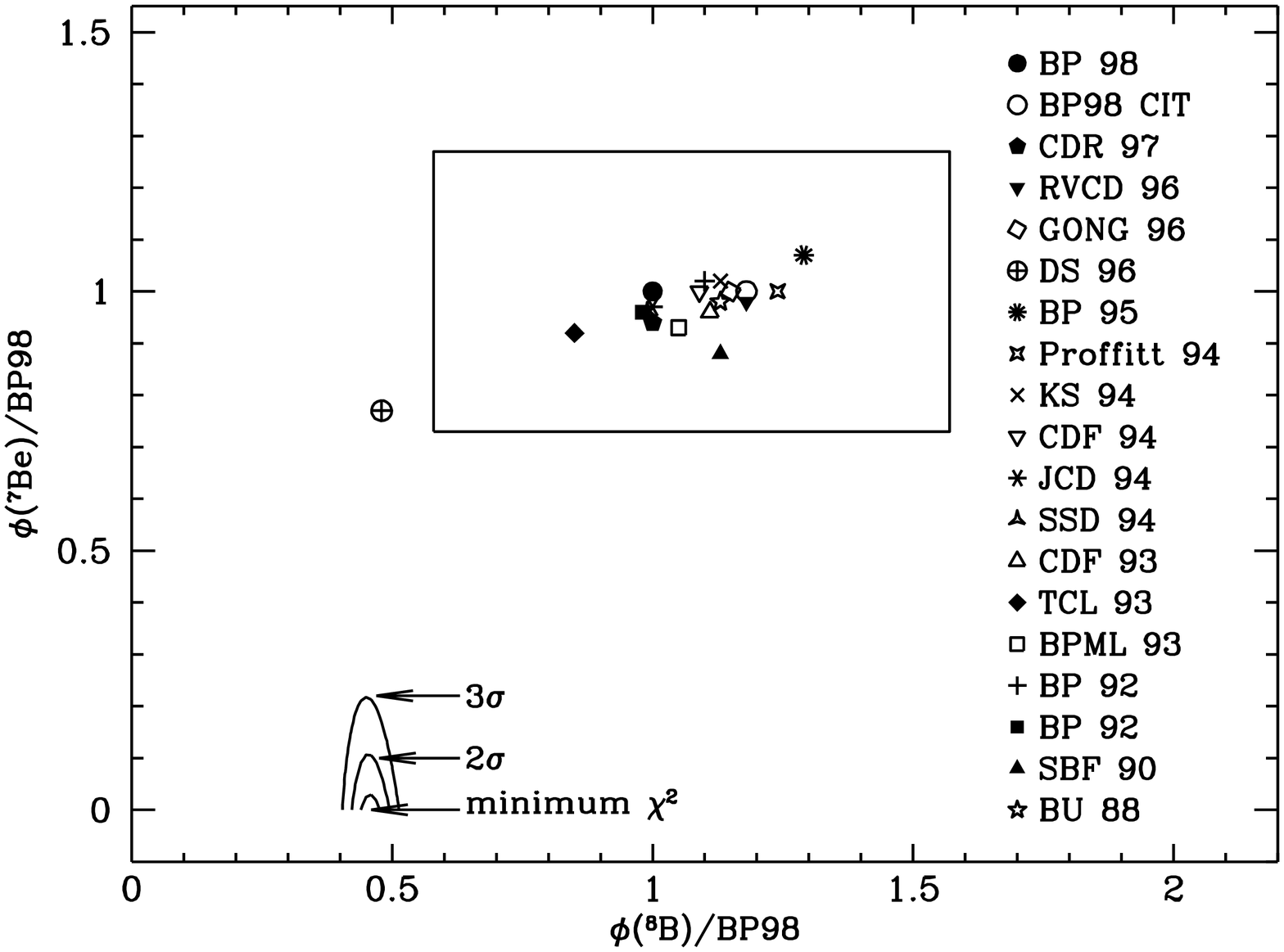,height=7cm,width=6cm}}}
\caption{SSM predictions, from ref.~\protect\cite{Bahcall98}}
\label{78}
\end{figure}
In \fig{78} one can see the predictions of various standard solar
models in the plane defined by the $^7$Be and $^8$B neutrino fluxes,
normalized to the predictions of the BP98 solar model~\cite{BP98}.
Abbreviations such as BP95, identify different solar models, as given
in ref.~\cite{models}.  The rectangular error box gives the $3\sigma$
error range of the BP98 fluxes. On the other hand the values of these
fluxes indicated by present data on neutrino event rates are shown by
the contours in the lower-left part of the figure. The best-fit $^7$Be
neutrino flux is negative!  The theoretical predictions clearly lie
well away from the $3\sigma$ contour, strongly suggesting the need for
new particle physics in order to account for the data~\cite{CF}.
Since possible non-standard astrophysical solutions are rather
constrained by helioseismology studies \cite{Bahcall98,helio97} one is
led to assume the existence of neutrino conversions, such as those
induced by very small neutrino masses.  Possibilities include the MSW
effect~\cite{MSW}, vacuum neutrino oscillations
\cite{Glashow:1987jj,bimax98} and, possibly, flavour changing neutrino
interactions \cite{Krastev:1997cp}. Moreover, if neutrinos have
transition magnetic moments then one may have, in addition, the
possibility of Majorana neutrino Spin-Flavour Precessions \cite{SFP}.
Based upon these there emerge two new solutions to the solar neutrino
problem: the Resonant~\cite{RSFP,akhmedov97} and the Aperiodic
Spin-Flavour Precession mechanisms \cite{Semikoz:1998xu}, based on
regular and random magnetic fields, respectively.

The recent 825--day data sample~\cite{sksol825} presents no major
surprises, except that the recoil energy spectrum produced by solar
neutrino interactions shows more events in the highest bins.  Barring
the possibly of poorly understood energy resolution effects, it has
been noted~\cite{bkhep} that if the flux for neutrinos coming from the
${\rm ^3He} ~+~ p \to {\rm ^4He} ~+~e^+ ~+~\nu_e $, the so-called
$hep$ reaction, is well above the (uncertain) SSM predictions, then
this could significantly influence the electron energy spectrum
produced by solar neutrino interactions in the high recoil region,
with hardly any effect at lower energies.
\begin{figure}
\centerline{\protect\hbox{
\psfig{file=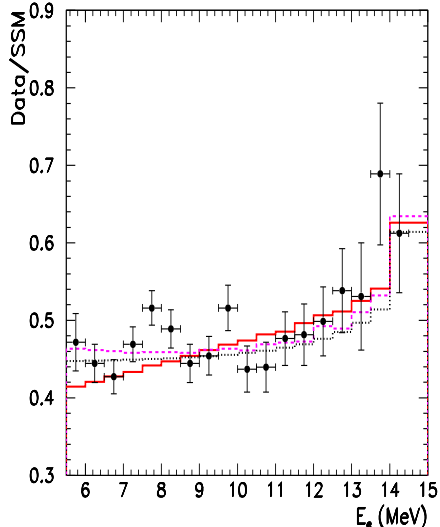,height=8cm,width=6.5cm}}}
\vglue -.8cm
\caption{Expected normalized recoil electron energy spectra versus
825--day Super--Kamiokande data, from~\protect\cite{valencia:1999}.  }
\label{spec800}
\end{figure}
Fig. \ref{spec800} shows the expected normalized recoil electron
energy spectrum compared with the most recent experimental
data~\protect\cite{sksol825}. The solid line represents the prediction
for the best--fit SMA solution with free $^8B$ and $hep$
normalizations (0.69 and 12 respectively), while the dotted line gives
the corresponding prediction for the best--fit LMA solution (1.15 and
34 respectively). Finally, the dashed line represents the prediction
for the best no-oscillation scheme with free $^8B$ and $hep$
normalizations (0.44 and 14, respectively). Clearly the spectra with
enhanced $hep$ neutrinos provide better fits to the data. However
Fiorentini et al~\cite{Fiorentini:1998xr} have argued that the
required $hep$ amount is too large to accept on theoretical grounds.
We look forward to the improvement of the situation.  The increasing
r\^ole played rate-independent observables such as the spectrum, as
well as seasonal and day-night asymmetries will eventually select
amongst different solutions of the solar neutrino problem.

\begin{figure}
\centerline{
\protect\hbox{\psfig{file=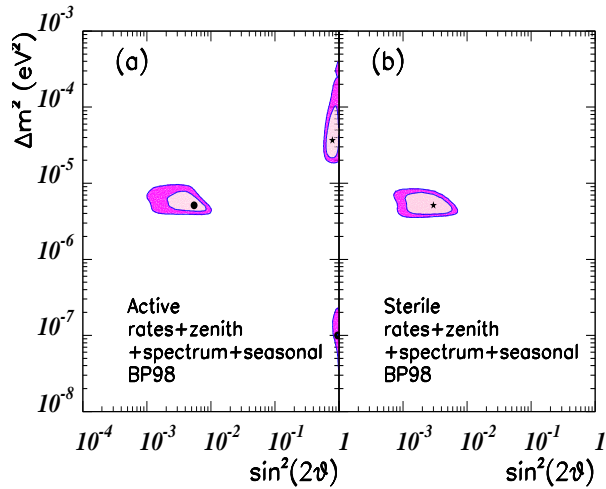,width=8.5cm,height=8cm}}}
\vglue -.8cm
\caption{Solar neutrino parameters  at 90~\% and 99~\% CL for 
  2-flavour MSW neutrino conversions with 825--day SK data sample in
  the BP98 model, from~\protect\cite{valencia:1999}.  }
\label{msw}
\end{figure}

The required solar neutrino parameters are determined through a
$\chi^2$ fit of the experimental data.  In~\fig{msw} we show the
allowed regions in $\Delta m^2$ and $\sin^2\theta$ from the
measurements of the total event rates at the Chlorine, Gallium and
Super--Kamiokande (825-day data sample) experiments, combined with the
zenith angle distribution, the recoil energy spectrum and the seasonal
dependence of the event rates, observed in Super--Kamiokande. Panels
{\bf(a)} and {\bf(b)} correspond to active-active and active-sterile
oscillations, respectively.  The best--fit points in each case are
indicated by a star~\cite{valencia:1999}, while the local best-fit
points are indicated by a dot.
An analysis with free $^8B$ and $hep$ normalizations has also been
given in ~\cite{valencia:1999} and does not change significantly the
allowed regions.

One notices from the analysis that rate-independent observables, such
as the electron recoil energy spectrum and the day-night asymmetry
(zenith angle distribution), are playing an increasing r\^ole in the
determination of solar neutrino parameters~\cite{valencia:1999}.  An
observable which has been neglected in most analyses of the MSW effect
and which could be sizeable in the large mixing angle regions (LMA and
LOW) is the seasonal dependence in the solar neutrino flux which would
result from the regeneration effect at the Earth and which has been
discussed in ref.~\cite{deHolanda:1999ty}.  This should play a more
significant r\^ole in future investigations.

\begin{figure}
\centerline{\protect\hbox{\psfig{file=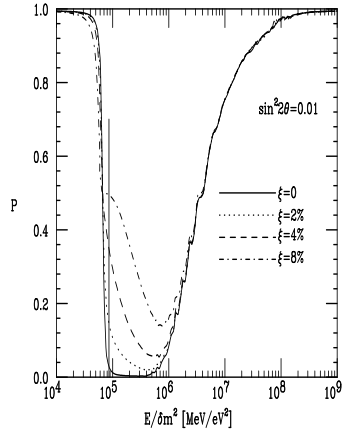,width=7cm,height=5.5cm,angle=90}}}
\vglue -.8cm
\caption{ Solar neutrino survival probability in a noisy Sun, 
from ref.~\protect\cite{noise}}
\label{Pnoise}
\end{figure}

A theoretical issue which has raised some interest recently is the
study of the possible effect of random fluctuations in the solar
matter density \cite{BalantekinLoreti,noise,noise2}. The possible
existence of such noise fluctuations at a few percent level is not
excluded by present helioseismology studies.
In \fig{Pnoise} we show averaged solar neutrino survival probability
as a function of $E/\Delta m^2$, for $\sin^2 2\theta = 0.01$. This
figure was obtained via a numerical integration of the MSW evolution
equation in the presence of noise, using the density profile in the
Sun from BP95 in ref.~\cite{models}, and assuming that the correlation
length $L_0$ (which corresponds to the scale of the fluctuation) is
$L_0 = 0.1 \lambda_m$, where $\lambda_m$ is the neutrino oscillation
length in matter. An important assumption in the analysis is that $
l_{free} \ll L_0 \ll \lambda_m$, where $l_{free} \sim 10 $ cm is the
mean free path of the electrons in the solar medium. The fluctuations
may strongly affect the $^7$Be neutrino component of the solar
neutrino spectrum so that the Borexino experiment should provide an
ideal test, if sufficiently small errors can be achieved. The
potential of Borexino in probing the level of solar matter density
fluctuations provides an additional motivation for the experiment
\cite{borexino}. 

The most popular alternative solution to the solar neutrino problem is
the {\sl vacuum oscillation solution} \cite{Glashow:1987jj} which
clearly requires large neutrino mixing and the adjustment of the
oscillation length so as to coincide roughly with the Earth-Sun
distance.
\begin{figure}
\centerline{\protect\hbox{\psfig{file=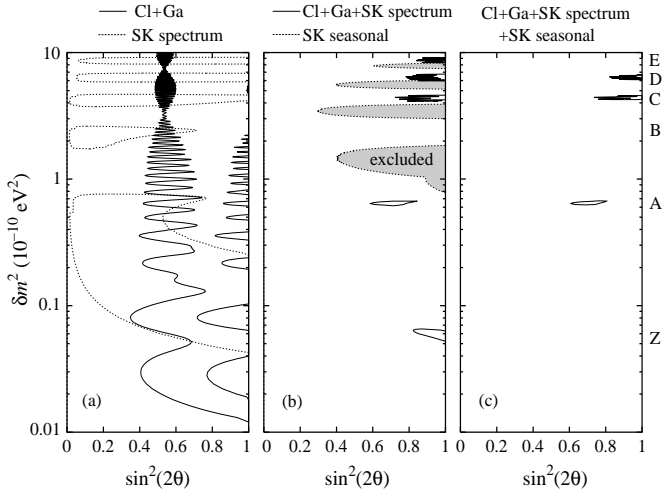,width=9cm,height=6.5cm}}}
\caption{Vacuum oscillation parameters, from ref.~\protect\cite{jso700}}
\label{vac99}
\end{figure}
Fig. \ref{vac99} shows the regions of just-so oscillation parameters
at the 95~\%~CL obtained in a recent fit of the data, including both
the rates, the recoil energy spectrum and seasonal effects, which are
expected in this scenario ~\cite{lisi} and could potentially help in
discriminating it from the MSW scenario.

\subsection{Atmospheric Neutrinos}
\vskip .1cm

Neutrinos produced as decay products in hadronic showers from cosmic
ray collisions with nuclei in the upper atmosphere have been observed
in several experiments
\cite{Super-Kamiokande,sk99,Frejus,Nusex,Kamiokande,IMB,Soudan}.
There has been a long-standing discrepancy between the predicted and
measured $\mu/e$ ratio of the muon ($\nu_\mu + \bar{\nu}_\mu$) over
the electron atmospheric neutrino flux ($\nu_e+\bar{\nu}_e$)
~\cite{atmreview}.  The anomaly has been found both in water Cerenkov
experiments (Kamiokande, Super-Kamiokande and IMB) as well as in the
iron calorimeter Soudan2 experiment.  Negative experiments, such as
Frejus and Nusex have much larger errors.
Although individual $\nu_\mu$ or $\nu_e$ fluxes are only known to
within $30\%$ accuracy, their ratio is predicted to within $5\%$ over
energies varying from 0.1~GeV to 100~GeV~\cite{atmfluxes}.  The most
important feature of the atmospheric neutrino
data~sample~\cite{atm52kt} is that it exhibits a {\sl
  zenith-angle-dependent} deficit of muon neutrinos.  Experimental
biases and uncertainties in the prediction of neutrino fluxes and
cross sections are unable to explain the data.

The most popular way to account for this anomaly is in terms of
neutrino oscillations. It has already been noted~\cite{atm98} that the
Chooz reactor data~\cite{Chooz} excludes the $\nu_{\mu} \to \nu_e$
channel, when all experiments are combined. So I concentrate here on
the other possible oscillation channels.

\begin{figure*}
\centerline{\protect\hbox{\epsfig{file=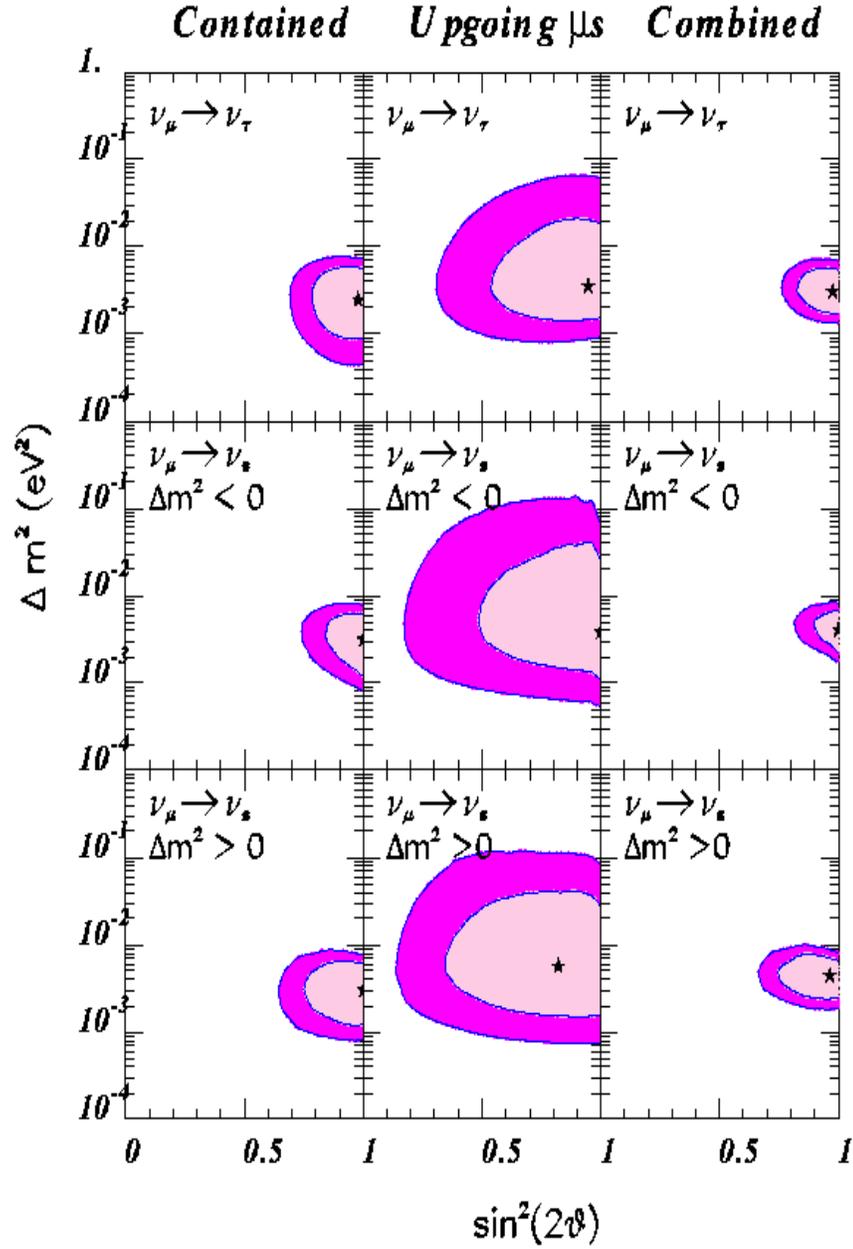,width=0.75\textwidth,
height=0.75\textheight}}}
\caption{
  Allowed regions of the oscillation parameters for various
  Super-Kamiokande data samples and oscillation channels, as labelled
  in the figure.  Best-fit points are denoted by a star in each case.}
\label{cont_osc}
\end{figure*}

The results of the most recent $\chi^2$ fit of the Super-Kamiokande
atmospheric neutrino data in the framework of the neutrino oscillation
hypothesis can be seen in Fig.~(\ref{cont_osc}), taken from
ref.~\cite{atmconcha}. This analysis updates previous studies in
ref.~\cite{atm98} and ~\cite{atmo98} and includes the upgoing muon
event samples.  This figure shows the allowed regions of oscillation
parameters at 90 and 99 \% CL.
Notice that matter effects lead to differences between the allowed
regions for the various channels.  For $\nu_\mu \to \nu_s$ with
$\Delta m^2>0$ matter effects enhance the oscillations for {\sl
  neutrinos} and therefore smaller values of the vacuum mixing angle
would lead to larger conversion probabilities, so that the regions are
larger than compared to the $\nu_\mu \to \nu_\tau$ case. For $\nu_\mu
\to \nu_s$ with $\Delta m^2<0$ the matter enhancement occurs only for
{\sl anti-neutrinos}, suppressing the conversion in $\nu_\mu$'s.
Since the yield of atmospheric neutrinos is bigger than that of
anti-neutrinos, clearly the matter effect suppresses the overall
conversion probability.  Therefore one needs in this case a larger
value of the vacuum mixing angle. This trend can indeed be seen by
comparing the regions in different columns of Fig.~(\ref{cont_osc}).

Notice that in all channels where matter effects play a r\^ole the
range of acceptable $\Delta m^2$ is slightly shifted towards larger
values, as compared with the $\nu_\mu \to \nu_\tau$ case. This follows
from the relation between mixing {\sl in vacuo} and in matter. In
fact, away from the resonance region, independently of the sign of the
matter potential, there is a suppression of the mixing inside the
Earth. As a result, the lower allowed $\Delta m^2$ value is higher
than for the $\nu_\mu \to \nu_\tau$ channel.

\begin{figure}
\centerline{\protect\hbox{\epsfig{file=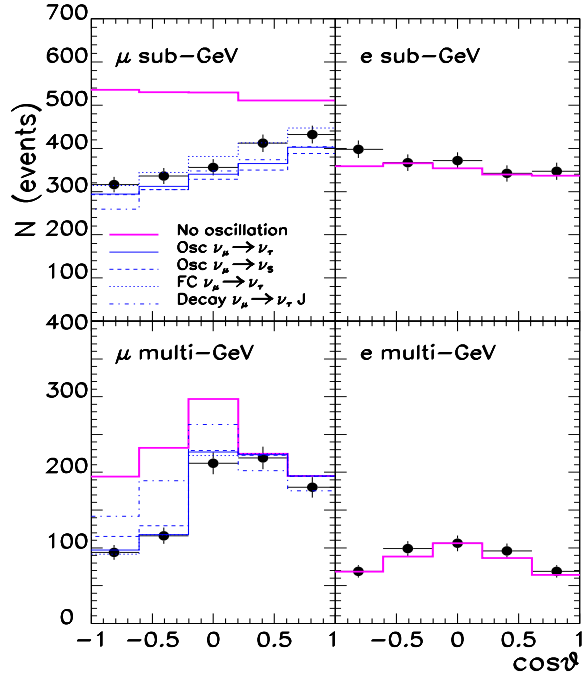,width=0.5\textwidth,height=9cm}}}
\caption{Angular distributions for  contained Super-Kamiokande 
  events, together with the SM prediction (no-oscillation) and the
  predictions for the best-fit points to the contained event data in
  various conversion mechanisms labelled in the figure. The error in
  the experimental points is only statistical.}
\label{ang_cont}  
\end{figure}

\begin{figure}
\centerline{\protect\hbox{\epsfig{file=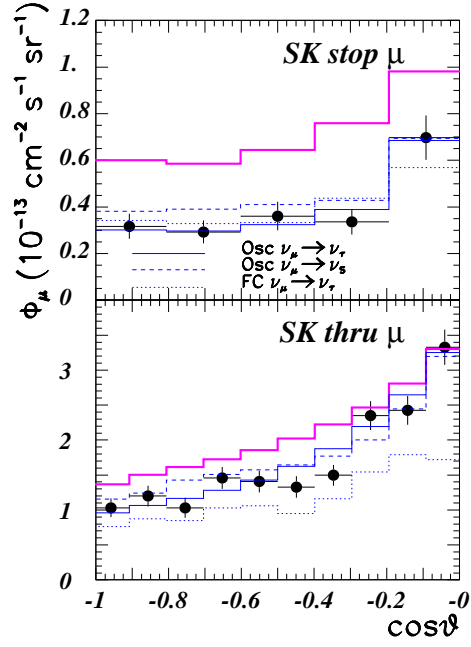,width=0.4\textwidth,height=9cm}}}
\caption{Angular distribution for Super-Kamiokande upgoing muon data
  together with the SM prediction (no-oscillation) as well as the
  prediction for the best-fit point to the full data sample for the
  different conversion mechanisms labelled in the figure. }
\label{ang_up}  
\end{figure}

Concerning the quality of the fits we note that the best-fit to the
full sample is obtained for the $\nu_\mu \to \nu_\tau$ channel,
although from the global analysis oscillations into sterile neutrinos
cannot be ruled out.  There is also an improvement in the quality of
the fits to the contained events as compared to previous analysis
performed with lower statistics \cite{atm98}. These features can be
easily understood by looking at the predicted zenith angle
distribution of the different event types for the various oscillation
channels shown in Fig.~(\ref{ang_cont}) and Fig.~(\ref{ang_up}). From
Fig.~(\ref{ang_cont}) one can see the excellent agreement between the
observed distributions of e-like events and the SM predictions.  This
has led to an improvement of the quality of the fit for any conversion
mechanism that only involves muons. From Fig.~(\ref{ang_up}) one can
also see that due to matter effects the distributions for upgoing
muons in the case of $\nu_\mu \to \nu_s$ are flatter than for $\nu_\mu
\to \nu_\tau$ \cite{lipari}.  The data show a somewhat steeper angular
dependence which is better described by $\nu_\mu \to \nu_\tau$
oscillations. In order to exploit this feature the Super-Kamiokande
collaboration has presented a preliminary partial analysis of the
angular dependence of the through-going muon data in combination with
the up-down asymmetry of partially contained events which seems indeed
to disfavour $\nu_\mu \to \nu_s$ oscillations at the 2--$\sigma$ level
\cite{sk99}.
For a comparison of the oscillation parameters as determined from the
atmospheric data with the sensitivity of the present accelerator and
reactor experiments, as well as the expectations of upcoming
long-baseline experiments see ref.~\cite{atm98}.

\subsection{LSND}
\vskip .1cm

\begin{figure}
\vglue -1cm
\centerline{\protect\hbox{\epsfig{file=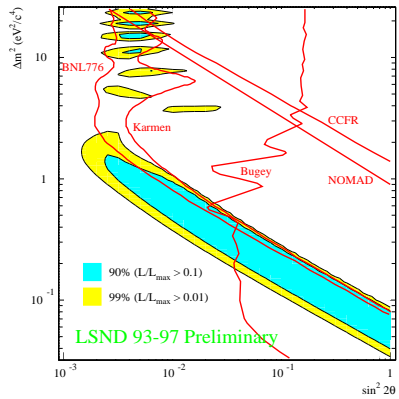,width=6cm,height=6cm}}}
\caption{Allowed LSND oscillation parameters versus competing
experiments~\protect\cite{louis} }
\label{darlsnd} 
\vglue -.5cm
\end{figure}

\begin{figure}
\centerline{\protect\hbox{\epsfig{file=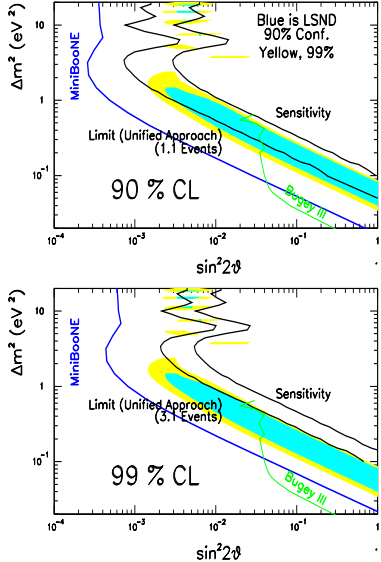,width=6cm,height=8cm}}}
\caption{Expected sensitivity of the proposed MiniBooNE
experiment~\protect\cite{louis} }
\label{miniboone} 
\end{figure}

The Los Alamos Meson Physics Facility looked for $\bar\nu_{\mu} \to
\bar\nu_{e}$ oscillations using $\bar\nu_\mu$ from $\mu^+$ decay at
rest \cite{LSND}. The $\bar\nu_e$'s are detected via the reaction
$\bar\nu_e\,p \to e^{+}\,n$, correlated with a $\gamma$ from $np \to d
\gamma$ ($2.2\,{\rm MeV}$).  The results indicate $\bar \nu_\mu \to
\bar \nu_e$ oscillations, with an oscillation probability of
($0.31^{+0.11}_{-0.10} \pm 0.05$)\%, leading to the oscillation
parameters shown in~\fig{darlsnd}.  The shaded regions are the
favoured likelihood regions given in ref.~\cite{LSND}.  The curves
show the 90~\% and 99~\% likelihood allowed ranges from LSND, and the
limits from BNL776, KARMEN1, Bugey, CCFR, and NOMAD.
A search for \nm $\to$ \ne oscillations has also been conducted by the
LSND collaboration.  Using \nm from $\pi^+$ decay in flight, the \ne
appearance is detected via the charged-current reaction $C(\ne,e^-)X$.
Two independent analyses are consistent with the above signature,
after taking into account the events expected from the \ne
contamination in the beam and the beam-off background.  If interpreted
as an oscillation signal, the observed oscillation probability of $2.6
\pm 1.0 \pm 0.5 \times 10^{-3}$, consistent with the evidence for
oscillation in the \bnm $\to$ \bne channel described above.
Fig.~\ref{miniboone} compares the LSND region with the expected
sensitivity from MiniBooNE, which was recently approved to run at
Fermilab~\cite{Louis:1998qf,louis}.
A possible confirmation of the LSND anomaly would be a discovery of
far-reaching implications.

\subsection{ Dark Matter}
\vskip .1cm

Galaxies as well as the large scale structure in the Universe should
arise from the gravitational collapse of fluctuations in the expanding
universe. They are sensitive to the nature of the cosmological dark
matter. The data on cosmic background temperature anisotropies on
large scales performed by the COBE satellite~\cite{cobe} combined with
cluster-cluster correlation data e.g. from IRAS~\cite{iras} can not be
reconciled with the simplest COBE-normalized $\Omega_m=1$ cold dark
matter (CDM) model, since it leads to too much power on small
scales. Adding to CDM neutrinos with mass of few eV (a scale similar
to the one indicated by the LSND experiment
\cite{LSND}) corresponding to $\Omega_\nu \approx 0.2$, results in an
improved fit to data on the nearby galaxy and cluster
distribution~\cite{cobe2}.  The resulting Cold + Hot Dark Matter
(CHDM) cosmological model is the most successful $\Omega_m=1$ model
for structure formation, preferred by inflation.  However, other
recent data have begun to indicate a lower value for $\Omega_m$, thus
weakening the cosmological evidence favouring neutrino mass of a few
eV in flat models with cosmological constant $\Omega_\Lambda = 1 -
\Omega_m$~\cite{cobe2}.  Future sky maps of the cosmic microwave 
background radiation (CMBR) with high precision at the MAP and PLANCK
missions should bring more light into the nature of the dark matter
and the possible r\^ole of neutrinos \cite{Raffelt:1999zg}.  Another
possibility is to consider unstable dark matter scenarios
\cite{Gelmini:1984pe}. For example, an MeV range tau neutrino may
provide a viable unstable dark matter scenario \cite{ma1} if the \nt
decays before the matter dominance epoch. Its decay products would add
energy to the radiation, thereby delaying the time at which the matter
and radiation contributions to the energy density of the universe
become equal. Such delay would allow one to reduce the density
fluctuations on the smaller scales purely within the standard cold
dark matter scenario.  Upcoming MAP and PLANCK missions may
place limits on neutrino stability \cite{Hannestad:1999xy} and rule
out such schemes.

\subsection{  Pulsar Velocities}
\vskip .1cm

One of the most challenging problems in modern astrophysics is to find
a consistent explanation for the high velocity of pulsars.
Observations \cite{veloc} show that these velocities range from zero
up to 900 km/s with a mean value of $450 \pm 50$ km/s.  An attractive
possibility is that pulsar motion arises from an asymmetric neutrino
emission during the supernova explosion. In fact, neutrinos carry more
than $99 \%$ of the new-born proto-neutron star's gravitational
binding energy so that even a $1 \%$ asymmetry in the neutrino
emission could generate the observed pulsar velocities.  This could in
principle arise from the interplay between the parity violation
present in weak interactions with the strong magnetic fields which are
expected during a SN explosion~\cite{Chugai,others}. However, it has
recently been noted~\cite{vilenkin98} that no asymmetry in neutrino
emission can be generated in thermal equilibrium, even in the presence
of parity violation. This suggests that an alternative mechanism is at
work.
Several neutrino conversion mechanisms in matter have been invoked as
a possible engine for powering pulsar motion.  They all rely on the
{\sl polarization} \cite{NSSV} of the SN medium induced by the strong
magnetic fields $10^{15}$ Gauss present during a SN explosion. This
would affect neutrino propagation properties giving rise to an angular
dependence of the matter-induced neutrino potentials. This would lead
in turn to a deformation of the "neutrino-sphere" for, say, tau
neutrinos and thus to an anisotropic neutrino emission.  As a
consequence, in the presence of non-vanishing $\nu_\tau$ mass and
mixing the resonance sphere for the $\nu_e-\nu_\tau$ conversions is
distorted.  If the resonance surface lies between the $\nu_\tau$ and
$\nu_e$ neutrino spheres, such a distortion would induce a temperature
anisotropy in the flux of the escaping tau-neutrinos produced by the
conversions, hence a recoil kick of the proto-neutron star.
This mechanism was realized in ref.~\cite{KusSeg96} invoking MSW
conversions \cite{MSW} with \mnt $\gsim$ 100 eV or so, assuming a
negligible $\nu_e$ mass. This is necessary in order for the resonance
surface to be located between the two neutrino-spheres.  It should be
noted, however, that such requirement is at odds with cosmological
bounds on neutrinos masses unless the $\tau$-neutrino is unstable.
On the other hand in ref.~\cite{ALS} a realization was proposed in the
resonant spin-flavour precession scheme (RSFP) \cite{RSFP}.  The
magnetic field would not only affect the medium properties, but would
also induce the spin-flavour precession through its coupling to the
neutrino transition magnetic moment \cite{SFP}.

Perhaps the simplest and probably most elegant suggestion was proposed
in ref.~\cite{Grasso:1998tt}, where the required pulsar velocities
would arise from anisotropic neutrino emission induced by resonant
conversions of massless neutrinos (hence no magnetic moment).
Raffelt and Janka~\cite{Janka:1999kb} have subsequently argued that
the asymmetric neutrino emission effect was overestimated in
~\cite{KusSeg96,ALS,Grasso:1998tt}, since the temperature variation
over the deformed neutrino-sphere is not an adequate measure for the
anisotropy of the neutrino emission. This would invalidate all
neutrino conversion mechanisms, leaving the pulsar velocity problem
without any viable solution. However Kusenko and Segr\'e still
maintain that sizeable pulsar kicks can arise from neutrino
conversions \cite{Kusenko:1998bk}. In any case invoking conversions
into sterile neutrinos could be an interesting possibility, since the
conversions could take place deeper in the star \cite{nuno98}.

\section{Making Sense of All That}
\vskip .1cm
 
Physics beyond the Standard Model is required in order to explain
solar and atmospheric neutrino data. While neutrino oscillations
provide an excellent fit and a powerful way to determine neutrino mass
and mixing, there is a plethora of alternative mechanisms, some of
which quite attractive, which could play an important r\^ole in the
interpretation of the data.  These include flavour changing neutrino
interactions both in the solar \cite{Krastev:1997cp} and atmospheric
~\cite{Gonzalez-Garcia:1998hj,fornengo99} neutrino problems,
Resonant~\cite{RSFP,akhmedov97} and the Aperiodic Spin-Flavour
Precession mechanisms \cite{Semikoz:1998xu} for solar neutrinos, which
use the transition magnetic moments of Majorana neutrinos \cite{SFP},
and the possibility of fast neutrino decays~\cite{V} which could play
a r\^ole in the atmospheric neutrino problem~\cite{atmdecay}. Moreover
I note that more exotic explanations of the undergound neutrino data
based upon violations of equivalence principle, Lorentz invariance and
CPT have been proposed~\cite{exotic}.  Nevertheless in what follows I
will assume the standard neutrino oscillation interpretation of the
data.

\subsection{Solar plus Atmospheric}
\vskip .1cm

These data can be accounted for with the three known neutrinos. They
fix the two mass splittings $\Delta {m_\odot^2}$ \& $\Delta
{m_{atm}^2}$, and two of the three neutrino mixing angles, the third
being small on account of the Chooz reactor results~\cite{Chooz}.
Such scenario can easily be accommodated in \emph{seesaw} theories of
neutrino mass since, in general, the mixing angles involved are not
predicted, in particular the maximal mixing indicated by the
atmospheric data and possibly also by the solar data can be
accomodated by hand.
In contrast, it is not easy to reconcile maximal or {\sl bi-maximal}
mixing of neutrinos~\cite{bimax98} with a predictive quark-lepton
\emph{unification seesaw} scheme that relates lepton and quark mixing
angles, since the latter are known to be small. For attempts to
reconcile solar and atmospheric data in unified models with specific
texture anzatze, see ref.~\cite{Lola:1998xp,Altarelli:1998ns}.

An alternative way to predict a hierarchical pattern of neutrino mass
and mixing, which naturally accomodates the possibility of maximal
mixing is to appeal to supersymmetry. In ref.~\cite{Romao:1999up} it
was shown that the simplest unified extension of the Minimal
Supersymmetric Standard Model with bi-linear R--Parity violation
provides a predictive scheme for neutrino masses which can account for
the observed atmospheric and solar neutrino anomalies in terms of
bi-maximal neutrino mixing. The maximality of the atmospheric mixing
angle arises dynamically, by minimizing the scalar potential of the
theory, while the solar neutrino problem can be accounted for either
by large or by small mixing oscillations. The spectrum is naturally
hierarchical, since only the tau neutrino picks up mass at the tree
level (though this may be itself calculable from renormalization-group
evolution from unification down to weak-sacle), by mixing with
neutralinos, while the masslessness of the other two neutrinos is
lifted only by calculable loop corrections.  Despite the smallness of
neutrino masses R-parity violation can be observable at present and
future high--energy colliders, providing an unambiguous cross-check of
the model, and the possibility of probing the neutrino anomalies at
accelerators.

Bi-maximal models may also be tested at the upcoming long-baseline
experiments or at a possible neutrino factory
experiment~\cite{Quigg:1999ap} through the CP violating phases, which
could lead to non-negligible CP asymmetries in neutrino
oscillations~\cite{CP_LBL_FAC}. Unfortunately the effects of the CP
violation intrinsic of the Majorana neutrino system ~\cite{2227} is
helicity suppressed ~\cite{Schechter:1981gk}, though a potential test
of the CP properties and Majorana nature of neutrinos has been
suggested in ref.  ~\cite{Pastor:1999iw}.

\subsection{Solar and Atmospheric  plus Dark Matter }
\vskip .1cm

The story gets more complicated if one wishes to account also for the
hot dark matter. The only possibility to fit the solar, atmospheric
and HDM scales in a world with just the three known neutrinos is if
all of them have nearly the same mass \cite{cm93}, of about $\sim$ 1.5
eV or so in order to provide the right amount of HDM \cite{cobe2} (all
three active neutrinos contribute to HDM). This can be arranged in a
unified \10 seesaw model where, to first approximation all neutrinos
lie at the above HDM mass scale ($\sim$ 1.5 eV), due to a suitable
horizontal symmetry, the splittings $\Delta {m_\odot^2}$ \& $\Delta
{m_{atm}^2}$ appearing as symmetry breaking effects. An interesting
fact is that the ratio $\Delta {m_\odot^2}/\Delta {m_{atm}^2}$ is
related to ${m_c}^2/{m_t}^2$~\cite{DEG}. There is no room in this case
to accommodate the LSND anomaly. To what extent this solution is
theoretically natural has been discussed recently in
ref.~\cite{Casas:1999ac}. One finds that the degeneracy is stable in
the phenomenologically relevant case where neutrinos have opposite CP
parities, leading to a suppression in the neutrino-less doble beta
decay rate ~\cite{Ma:1999ut}.

\subsection{Solar \& Atmospheric with Dark Matter \& LSND: Four-Neutrino Models}
\vskip .1cm

An alternative way to include hot dark matter scale is to invoke a
fourth light sterile neutrino~\cite{ptv92,pv93,cm93}. As a bonus we
can accomodate the LSND hint. The sterile neutrino \ns must also be
light enough in order to participate in the oscillations together with
the three active neutrinos. Since it is an \21 singlet it does not
affect the invisible Z decay width, well-measured at LEP.  The
theoretical requirements are:
\bi 
\item
to understand what keeps the sterile neutrino light, since the \21
gauge symmetry would allow it to have a large bare mass
\item
to account for the maximal neutrino mixing indicated by the
atmospheric data, and possibly by the solar
\item to account from first principles for the scales $\Delta
  {m_\odot^2}$, $\Delta {m_{atm}^2}$ and $\Delta m_{LSND/HDM}^2$ 
\ei
With this in mind we have formulated the simplest maximally symmetric
schemes, denoted as $(e\tau)(\mu~s)$~\cite{ptv92} and $(es)(\mu\tau)$
~\cite{pv93}, respectively. One should realize that a given scheme
(mainly the structure of the leptonic charged current) may be realized
in more than one theoretical model. For example, an alternative to the
model in ~\cite{pv93} was suggested in ref.~\cite{cm93}. Higher
dimensional theories contain light sterile neutrinos which can arise
from the bulk sector and reproduce the basic features of these
models~\cite{nubrane}. For a recent discussion of the experimental
constraints on four-neutrino mixing see ref.~\cite{Giunti:1999hb}. For
alternative theoretical and phenomenological scenarios see
ref.~\cite{ptvlate,smir}.
 
\begin{figure}[t]
\centerline{\protect\hbox{\psfig{file=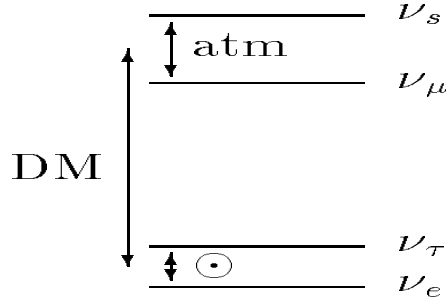,width=6cm,height=4cm}}}
\caption{$(e\tau)(\mu~s)$ scheme: \ne- \nt conversions explain the
solar neutrino data and \nm- \ns oscillations account for the
atmospheric deficit, ref.~\protect\cite{ptv92}.}
\label{ptv}
\vglue -.2cm
\end{figure}

\begin{figure}
\centerline{\protect\hbox{\psfig{file=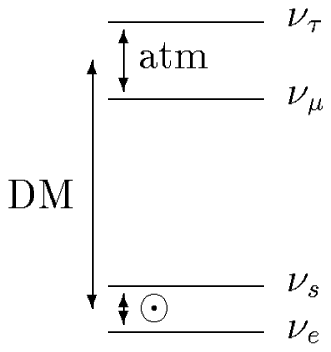,width=6cm,height=4cm}}}
\caption{$(es)(\mu\tau)$ scheme: \ne- \ns conversions explain the
solar neutrino data and \nm- \nt oscillations account for the
atmospheric deficit, ref.~\protect\cite{pv93}.}
\label{pv}
\vglue -.2cm
\end{figure}

Although many of the phenomenological features arise also in other
models, here I concentrate the discussion mainly on the theories
developed in ref.~\cite{ptv92,pv93}. These are characterized by a very
symmetric mass spectrum in which there are two ultra-light neutrinos
at the solar neutrino scale and two maximally mixed almost degenerate
eV-mass neutrinos (the LSND/HDM scale~\cite{pvhdm}), split by the
atmospheric neutrino scale~\cite{ptv92,pv93}. 
Before the global U(1) lepton symmetry breaks the heaviest neutrinos
are exactly degenerate, while the other two are massless
\cite{OLDsterilemodel}.  After the U(1) breaks down the heavier
neutrinos split and the lighter ones get mass.
The scale $\Delta m^2_{LSND/HDM}$ is generated radiatively at
one--loop due to the additional Higgs bosons, while the splittings
$\Delta m^2_{atm}$ and $\Delta {m^2}_\odot$ are two--loop effects.
The models are based only on weak-scale physics: no large mass scale
is introduced.  They {\sl explain} the lightness of the sterile
neutrino
\footnote{In higher dimensional theries such sterile neutrinos may
  arise from bulk matter and be light without need for a protecting
  symmetry, see ref.~\protect\cite{nubrane}. },
the large mixing required by the atmospheric neutrino data, as well as
the generation of the mass splittings responsible for solar and
atmospheric neutrino conversions as natural consequences of the
underlying lepton--number symmetry and its breaking.  They are minimal
in the sense that they add a single \21 singlet lepton to the SM.  The
models differ according to whether the \ns lies at the dark matter
scale or at the solar neutrino scale. In the $(e\tau)(\mu~s)$ scheme
the \ns lies at the LSND/HDM scale, as illustrated in \fig{ptv}
while in the alternative $(es)(\mu\tau)$ model, \ns is at the solar
\neu scale as shown in \fig{pv} \cite{pv93}.
In the $(e\tau)(\mu~s)$ case the atmospheric neutrino puzzle is explained
by \nm to \ns oscillations, while in $(es)(\mu\tau)$ it is due to
 \nm to \nt oscillations. Correspondingly, the deficit of solar
\neus is explained in the first case by \ne to \nt conversions, while
in the second the relevant channel is \ne to $\nu_s$.  
In both models ~\cite{ptv92,pv93} one predicts close-to-maximal mixing
in the atmospheric neutrino sector, a feature which emerges as the
global best--fit points in the analyses discussed above.

The presence of additional weakly interacting light particles, such as
our light sterile neutrino, is constrained by BBN since the \ns would
enter into equilibrium with the active neutrinos in the early Universe
(and therefore would contribute to $N_\nu^{max}$) via neutrino
oscillations~\cite{bbnsterile}, unless $\Delta m^2 sin^42\theta \lsim
3\times 10^{-6}~~eV^2$.  Here $\Delta m^2$ denotes a typical
mass-square difference of the active and sterile species and $\theta$
is the vacuum mixing angle.  However, systematic uncertainties in the
BBN bounds still caution us not to take them too literally. For
example, it has been argued that present observations of primordial
Helium and deuterium abundances may allow up to $N_\nu = 4.5$ neutrino
species if the baryon to photon ratio is small~\cite{sarkar}. Adopting
this as a limit, clearly both models described above are consistent.
Should the BBN constraints get tighter \cite{Fiorentini:1998fv} e.g.
$N_\nu^{max} < 3.5$ they could rule out the $(e\tau)(\mu~s)$ model,
and leave out only the competing scheme as a viable alternative.
However the possible r\^ole of a primordial lepton asymmetry might
invalidate this conclusion, for recent work on this see
ref.~\cite{Foot:1997qc}.

It is well-known that the neutral-to-charged current ratios are
important observables in neutrino oscillation phenomenology,
especially sensitive to the existence of singlet neutrinos, light or
heavy \cite{2227}.  On this basis the two models above would be
distinguishable at future neutral-current-sensitive solar and
atmospheric neutrino experiments.  For example they may be tested in
the SNO experiment~\cite{SNO} once they measure the solar neutrino
flux ($\Phi^{NC}_{\nu}$) in their neutral current data and compare it
with the corresponding CC value ($\Phi^{CC}_{\nu}$).  If the solar
neutrinos convert to active neutrinos, as in the $(e\tau)(\mu~s)$
model, then one expects $\Phi^{CC}_{\nu}/\Phi^{NC}_{\nu}$ around 0.5,
whereas in the $(es)(\mu\tau)$ scheme (\ne conversion to \ns), the
above ratio would be nearly $ \simeq 1$.  
Looking at pion production via the neutral current reaction
$\nu_{\tau} + N \to \nu_{\tau} +\pi^0 +N$ in the atmospheric data
might also help in distinguishing between these two
possibilities~\cite{vissani}, since this reaction is absent in the
case of sterile neutrinos, but would exist in the $(es)(\mu\tau)$
scheme.

If light sterile neutrinos indeed exist one can show that they might
contribute to a cosmic hot dark matter component and to an increased
radiation content at the epoch of matter-radiation equality. These
effects leave their imprint in sky maps of the cosmic microwave
background radiation (CMBR) and may thus be detectable with the very
high precision measurements expected at the upcoming MAP and PLANCK
missions as noted in ref.~\cite{Raffelt:1999zg}.

\subsection{Heavy Tau Neutrino}
\vskip .1cm

Finally, the door is not closed to heavy neutrinos. Indeed, an
alternative to the inclusion of hot dark matter is to simulate its
effects through the late decay of an MeV tau neutrino~\cite{ma1}, in
the presence of a light sterile neutrino.  Indeed such a model was
presented \cite{JV95} where an unstable MeV Majorana tau neutrino
naturally reconciles the cosmological observations of large and
small-scale density fluctuations with the cold dark matter picture.
The model assumes the spontaneous violation of a global lepton number
symmetry at the weak scale.  The breaking of this symmetry generates
the cosmologically required decay of the \nt with lifetime
$\tau_{\nu_\tau} \sim 10^2 - 10^4$ sec, as well as the masses and
oscillations of the three light \neus \ne, \nm and $\nu_s$ which may
account for the present solar and atmospheric data, though a dedicated
three-neutrino fit in which one of the neutrinos is sterile would be
desirable.  

\section{In conclusion}
\vskip .1cm

The angle-dependent atmospheric neutrino deficit provides, together
with the solar neutrino data, a strong evidence for physics beyond the
Standard Model. Small neutrino masses provide the simplest, but not
unique, explanation of the data.  Allowing for alternative
explanations of the underground experiments involving non-standard
neutrinos opens new possibilities involving either massless or even
very heavy cosmologically unstable neutrinos, which naturally arise in
many models. From this point of view, it is still too early to infer
with great certainty neutrino masses and angles from underground
experiments alone. Keeping within the framework of the standard
neutrino oscillation interpretation of the data, one has an
interesting possibility of bi-maximal neutrino mixing and of testing
the neutrino anomalies not only at the upcoming long-baseline or
neutrino factory experiments, but also at high--energy accelerators.
On the other hand if the LSND result stands the test of time, this
would be a strong indication for the existence of a light sterile
neutrino.
The two most attractive ways to reconcile underground observations
with LSND invoke either \ne- \nt conversions to explain the solar
data, with \nm- \ns oscillations accounting for the atmospheric
deficit, or the opposite. At the moment the latter is favored by the
atmospheric data.  These two basic schemes have distinct implications
at future neutral-current-sensitive solar \& atmospheric neutrino
experiments, such as SNO and Super-Kamiokande.  To end up on a
phylosophical mood I would say that it is important to search for
manifestations of massive and/or non-standard neutrinos at the
laboratory in an unbiased way. Though most of the recent excitement
now comes from underground experiments, one should bear in mind that
models of neutrino mass may lead to a plethora of new signatures which
may be accessible also at high--energy accelerators, thus illustrating
the complementarity between the two approaches.

\vskip .3cm

I am grateful to the Organizers for the kind hospitality and to all my
collaborators, especially Concha Gonzalez-Garcia and her student
Carlos Pe\~na for the re-analysis of solar neutrino data. This work
was supported by DGICYT grant PB95-1077 and by the TMR contract
ERBFMRX-CT96-0090.


\vskip .7cm

\small
\begin{thebibliography}{99}
\baselineskip=.437cm

\bibitem{Cl99} B.T.~Cleveland {\it et al}\/., {\sl Astrophys. J.}\/ {\bf
    496}, 505 (1998).
       
\bibitem{Ga99} GALLEX Collaboration (W.~Hampel {\it et al}\/.), {\sl
    Phys. Lett.}\/ {\bf B447}, 127 (1999); T.~Kirsten, Talk at the
  Sixth international workshop on topics in astroparticle and
  underground physics September, TAUP99, Paris, September 1999; SAGE
  Collaboration (J.N.~Abdurashitov {\it et al}\/.), {\sl Phys. Rev.}\/
  {\bf C 59}, 2246 (1999); SAGE Collaboration (J.N.~Abdurashitov {\it
    et al}.), astro-ph/9907113.

\bibitem{sksol504}
Y.~Fukuda {\it et al.}  [Super-Kamiokande Collaboration],
Phys. Rev. Lett. {\bf 82}, 1810 (1999) hep-ex/9812009;
Super--Kamiokande Collaboration, Y. Fukuda et al., Phys. Rev. Lett.
{\bf 82}, 2430 (1999).

\bibitem{sksol825} 
Y. Suzuki, talk a the ``XIX International Symposium on
Lepton and Photon Interactions at High Energies'', Stanford University, 
August 9-14, 1999.

\bibitem{atmexpold} NUSEX Collaboration, M. Aglietta {\sl et al.},
Europhys.  Lett.  {\bf 8}, 611 (1989); Fr\'ejus Collaboration, Ch.
Berger {\sl et al.}, Phys.  Lett.  {\bf B227}, 489 (1989); IMB
Collaboration, D. Casper {\sl et al.}, Phys. Rev. Lett.  {\bf 66},
2561 (1991); R. Becker-Szendy {\sl et al.}, Phys. Rev. {\bf D46}, 3720
(1992); Kamiokande Collaboration, H. S. Hirata {\sl et al.},
Phys. Lett. {\bf B205}, 416 (1988) and Phys. Lett. {\bf B280}, 146
(1992); Kamiokande Collaboration, Y. Fukuda {\sl et al.}, Phys.
Lett. {\bf B335}, 237 (1994); Soudan Collaboration, W.  W.  M Allison
{\sl et al.}, Phys.  Lett.  {\bf B391}, 491 (1997)

\bibitem{atm52kt} Y.~Fukuda {\it et al.}  [Super-Kamiokande
  Collaboration], ``Evidence for oscillation of atmospheric
  neutrinos," Phys. Rev. Lett. {\bf 81}, 1562 (1998) hep-ex/9807003;
Y. Fukuda et al., 
 Phys. Rev. Lett. {\bf 82}, 2644 (1998), hep-ex/9812014 and
 Y.~Fukuda {\it et al.}
 hep-ex/9908049.  The most recent 52 kton-yr data were presented in
 the invited talk by A. Mann at the XIX Lepton Photon Symposium at
 Stanford University, August 1999.

\bibitem{fae} For a review see J. W. F. Valle, {\it Gauge Theories and
the Physics of Neutrino Mass}, \ppnp{26}{91}{91-171}

\bibitem{Schechter:1982bd}
J.~Schechter and J.W.~Valle, Phys. Rev. {\bf D25}, 2951 (1982); for
reviews see A.~Morales, ``Review on double beta decay experiments and
comparison with theory," hep-ph/9809540

\bibitem{Schechter:1981gk}
J.~Schechter and J.W.~Valle,
Phys.\ Rev.\ {\bf D23} (1981) 1666.

\bibitem{MSW}
A.Y.~Smirnov and S.P.~Mikheev, ``Neutrino Oscillations In Matter With
Varying Density," {\it In *Tignes 1986, Proceedings, '86 massive
neutrinos* 355-372};  L. Wolfenstein, \pr{D20}{79}{2634}.

\bibitem{valencia:1999}
M.C.~Gonzalez-Garcia, P.C.~de Holanda, C.~Pena-Garay, and
J.~W.~F.~Valle, {\sl Status of the MSW Solutions of the Solar Neutrino
Problem}, hep-ph/9906469, to be published in  Nucl. Phys. {\bf B}

\bibitem{Valle:1987gv}
J.W.~Valle,
Phys.\ Lett.\ {\bf 199B} (1987) 432.

\bibitem{Guzzo:1991cp}
M.M.~Guzzo and S.T.~Petcov,
Phys.\ Lett.\ {\bf B271} (1991) 172.

\bibitem{Nunokawa:1996tg}
H.~Nunokawa, Y.Z.~Qian, A.~Rossi and J.W.~Valle,
Phys.\ Rev.\ {\bf D54} (1996) 4356
hep-ph/9605301.

\bibitem{Krastev:1997cp}
P.I.~Krastev and J.N.~Bahcall, ``FCNC solutions to the solar neutrino
problem," hep-ph/9703267.

\bibitem{Gonzalez-Garcia:1998hj}
M.C.~Gonzalez-Garcia {\it et al.}, Phys. Rev. Lett. {\bf 82} (1999)
3202, hep-ph/9809531.

\bibitem{WYLER}
D. Wyler, L. Wolfenstein, \np{B218}{83}{205}

\bibitem{SST}
R. Mohapatra, J. W. F. Valle, \pr {D34} {86} {1642};
J. W. F. Valle, \nps{B11} {89} {118-177}

\bibitem{BER}
J. Bernabeu, A. Santamaria, J. Vidal, A. Mendez, J. W. F. Valle, 
\pl {B187} {87} {303}; J. G. Korner, A. Pilaftsis, K. Schilcher,
\pl {B300} {93} {381}

\bibitem{3E}
M. C. Gonzalez-Garcia, J. W. F. Valle, \mpl{A7}{92}{477};
erratum \mpl{A9}{94}{2569}; A. Ilakovac, A. Pilaftsis, 
\np{B437}{95}{491}; A. Pilaftsis, \mpl{A9}{94}{3595}

\bibitem{Hall:1986dx}
L.J.~Hall, V.A.~Kostelecky and S.~Raby, Nucl. Phys. {\bf B267}, 415
(1986).

\bibitem{V} 
J. W. F. Valle, \pl {B131} {83}{87};
G. Gelmini, J. W. F. Valle, \pl {B142} {84}{181};
J. W. F. Valle, \pl {B159} {85}{49};
A. Joshipura, S. Rindani, \pr{D46}{92}{3000}

\bibitem{SFP} J.~Schechter, J.~W.~F. Valle, Phys.~Rev.~{\bf D24}
1883,~(1981), Err. ibid.{\bf D25}~283,~(1982).

\bibitem{RSFP}     
E.Kh. Akhmedov, \pl{ B213}{88}{64-68}; C. S. Lim and W. Marciano,
\pr{D37}{88}{1368}

\bibitem{akhmedov97} E.Kh.~Akhmedov, {\em The neutrino magnetic moment
and time variations of the solar neutrino flux}, hep-ph/9705451

\bibitem{Semikoz:1998xu}
V.B.~Semikoz, A.A.~Bykov, V.Y.~Popov, A.I.~Rez and D.D.~Sokoloff,
hep-ph/9808274.

\bibitem{LSND} C. Athanassopoulos, [LSND Collaboration], 
\prl{75}{95}{2650}; \prl{ 77}{96} {3082 }; 
C. Athanassopoulos et al,  Phys. Rev. Lett. {\bf 81}, 1774 (1998)

\bibitem{Louis:1998qf} W.C.~Louis [LSND Collaboration],
Prog. Part. Nucl. Phys. {\bf 40}, 151 (1998).

\bibitem{ptv92}
J.~T. Peltoniemi, D.~Tommasini, and J~W~F Valle,
\pl {B298}{93}{383}

\bibitem{pv93}
J.~T. Peltoniemi, and J~W~F Valle, \np{B406}{93}{409}

\bibitem{cm93}
D.O.~Caldwell and R.N.~Mohapatra, \pr{D48}{93}{3259}

\bibitem{Raffelt:1999zg}
S.~Hannestad and G.~Raffelt,  Phys. Rev. {\bf D59}, 043001
(1999) astro-ph/9805223.

\bibitem{Bahcall98} J.~N. Bahcall, astro-ph/9808162

\bibitem{BP98} J. N. Bahcall, S. Basu and M. H. Pinsonneault,
Phys. Lett. B 433 (1998) 1.

\bibitem{models} (GONG) J. Christensen-Dalsgaard et al., GONG
Collaboration, Science 272 (1996) 1286; (BP95) J. N. Bahcall and
M. H. Pinsonneault, Rev. Mod. Phys.  67 (1995) 781; (KS94)
A. Kovetz and G. Shaviv, Astrophys. J. 426 (1994) 787; (CDF94)
V. Castellani, S. Degl'Innocenti, G. Fiorentini, L.M. Lissia and
B. Ricci, Phys. Lett. B 324 (1994) 425; (JCD94)
J. Christensen-Dalsgaard, Europhys. News 25 (1994) 71; (SSD94)
X. Shi, D.N. Schramm and D.S.P. Dearborn, Phys. Rev. D 50 (1994) 2414;
(DS96) A. Dar and G. Shaviv, Astrophys. J. 468 (1996) 933;
(CDF93) V. Castellani, S. Degl'Innocenti and G. Fiorentini, Astron.
Astrophys. 271 (1993) 601; (TCL93) S. Turck-Chi\`eze and I. Lopes,
Astrophys. J. 408 (1993) 347; (BPML93) G. Berthomieu, J. Provost,
P.  Morel and Y. Lebreton, Astron. Astrophys. 268 (1993) 775;
(BP92) J.N. Bahcall and M.H. Pinsonneault, Rev. Mod. Phys. 64 (1992)
885; (SBF90) I.-J. Sackman, A.I. Boothroyd and W.A. Fowler,
Astrophys. J. 360 (1990) 727; (BU88) J.N. Bahcall and R.K. Ulrich,
Rev. Mod. Phys. 60 (1988) 297; (RVCD96) O. Richard, S. Vauclair,
C. Charbonnel and W.A. Dziembowski, Astron. Astrophys. 312 (1996)
1000; (CDR97) F. Ciacio, S. Degl'Innocenti and B. Ricci,
Astron. Astrophys. Suppl. Ser.  123 (1997) 449.

\bibitem{CF}
J. N. Bahcall, \pl{B338}{94}{276};
V. Castellani, {\it et al} \pl{B324}{94}{245};
N. Hata, S. Bludman, and P. Langacker, \pr{D49}{94}{3622};
V. Berezinsky, {\rm Comm. on Nucl. and Part. Phys.} {\bf 21} 
(1994) 249

\bibitem{helio97} J.N.~Bahcall, M.H.~Pinsonneault, S.~Basu and
J.~Christensen-Dalsgaard, Phys.~Rev.~Lett. 78 (1997) 171. 

\bibitem{Glashow:1987jj} V.~Barger, K.~Whisnant and R.J.~Phillips,
  Phys. Rev. {\bf D24}, 538 (1981); S.L.~Glashow and L.M.~Krauss,
  Phys. Lett. {\bf 190B}, 199 (1987); S.L.~Glashow, P.J.~Kernan and
  L.M.~Krauss, Phys. Lett. {\bf B445}, 412 (1999)
  
\bibitem{bimax98} V.~Barger, S.~Pakvasa, T.J.~Weiler and K.~Whisnant,
  Phys. Lett. {\bf B437}, 107 (1998), hep-ph/9806387;
  E.~Torrente-Lujan, Phys.\ Lett.\ {\bf B389} (1996) 557; S.~Davidson
  and S.F.~King, Phys.  Lett. {\bf B445}, 191 (1998).

\bibitem{bkhep}
J.N.~Bahcall and P.I.~Krastev, Phys. Lett. {\bf B436}, 243 (1998);
R.~Escribano, J.M.~Frere, A.~Gevaert and D.~Monderen, Phys. Lett. {\bf
B444}, 397 (1998).

\bibitem{Fiorentini:1998xr}
G.~Fiorentini, V.~Berezinsky, S.~Degl'Innocenti and B.~Ricci, ``Bounds
on hep neutrinos," Phys. Lett. {\bf B444}, 387 (1998), astro-ph/9810083.

\bibitem{deHolanda:1999ty} P.C.~de Holanda, C.~Pena-Garay,
  M.C.~Gonzalez-Garcia and J.~W.~F.~Valle, ``Seasonal dependence in
  the solar neutrino flux," Phys.\ Rev.\ {\bf D60} (1999) 093010,
  hep-ph/9903473; see also J.N.~Bahcall, P.I.~Krastev and
  A.Y.~Smirnov, hep-ph/9905220.

\bibitem{BalantekinLoreti}
A.B. Balantekin, J.M. Fetter and F.N. Loreti, \pr{D54}{96}{3941-3951};
F. N. Loreti and A. B. Balantekin, Phys. Rev. {\bf D50} (1994) 4762; 
F. N. Loreti {\it et al.}, Phys. Rev. {\bf D52} (1996) 6664. 

\bibitem{noise}
H. Nunokawa, A. Rossi, V. Semikoz, J. W. F. Valle, 
\np{B472}{96}{495-517} [see also talk at Neutrino 96, 
hep-ph/9610526]

\bibitem{noise2} 
P.~Bamert, C.P.~Burgess and D.~Michaud, Nucl. Phys. {\bf B513}, 319
(1998); C.P. Burgess, hep-ph/9711425; C.P.~Burgess and D.~Michaud,
Annals Phys. {\bf 256}, 1 (1997) and hep-ph/9611368.

\bibitem{borexino} C. Arpesella at al., Proposal of the Borexino
experiment (1991).

\bibitem{jso700} V. Barger, K. Whisnant, hep-ph/9903262; S.~Goswami,
  D.~Majumdar and A.~Raychaudhuri,
  hep-ph/9909453.

\bibitem{lisi} 
S.P. Mikheyev, A.Yu. Smirnov  Phys.Lett. {\bf B429} (1998) 343-348;
B. Faid, G. L. Fogli, E. Lisi, D. Montanino, hep-ph/9805293

\bibitem{Super-Kamiokande} 
Y. Fukuda {\it et al.}, Phys. Lett.  {\bf B433}, 9 (1998);
Phys. Lett.\ {\bf B436}, 33 (1998)

\bibitem{sk99} M. Nakahata, talk at {\sl Sixth International Workshop
    on Topics in Astroparticle and Underground Physics}, TAUP99, Paris
  September 1999.

\bibitem{Frejus}  
K. Daum {\it et al.} Z. Phys. {\bf C66}, 417 (1995).

\bibitem{Nusex} 
M. Aglietta {\it et al.}, Europhys. Lett. {\bf 8}, 611 (1989).

\bibitem{Kamiokande} 
H. S. Hirata {\it et al.}, Phys. Lett. {\bf B280}, 146 (1992); 
Y. Fukuda {\it et al.}, {\em ibid} {\bf B335}, 237 (1994).

\bibitem{IMB} 
R. Becker-Szendy {\it et al.}, Phys. Rev. {\bf D46}, 3720 (1992).

\bibitem{Soudan} 
W. W. M. Allison {\it et al.}, Phys.\ Lett.\ {\bf B449}, 137 (1999).

\bibitem{atmreview} 
T.~K. Gaisser, F. Halzen and T. Stanev, Phys. Rep. {\bf 258}, 174
(1995)

\bibitem{atmfluxes} T. K. Gaisser and T. Stanev, Phys. Rev.  {\bf D57}
  1977 (1998); G.\,Barr, T.\,K.\,Gaisser and T.\,Stanev, Phys. Rev.
  {\bf D\,39} (1989) 3532 and Phys.\ Rev.\ {\bf D38}, 85; V. Agrawal
  {\it et al.}, Phys. Rev. {\bf D53}, 1314 (1996); L. V. Volkova, Sov.
  J. Nucl. Phys. {\bf 31}, 784 (1980); M. Honda, T. Kajita, S.
  Midorikawa and K. Kasahara, Phys.\ Rev.\ {\bf D52}, 4985 (1995).

\bibitem{atm98} M.~C. Gonzalez-Garcia, H. Nunokawa, O.~L.~G. Peres,
T. Stanev, J.~W.~F. Valle, \pr{D58}{98}{033004} [hep-ph 9801368]; for
the 535~days~data~sample update, and the comparison of active and
sterile channels see M.C.~Gonzalez-Garcia, H.~Nunokawa, O.L.~Peres and
J.~W.~F.~Valle, Nucl. Phys. {\bf B543}, 3 (1998), hep-ph/9807305.

\bibitem{Chooz}
M. Appollonio et al.
Phys. Lett. {\bf B420} 397(1998), hep-ex/9711002

\bibitem{atmconcha}
M.~C. Gonzalez-Garcia, talk at International Workshop on Particles in
Astrophysics and Cosmology: From Theory to Observation, Valencia,
Spain, May 3-8, 1999, To be published in Nucl. Phys. B (Proc. Suppl.),
Ed. V. Berezinsky, G. Raffelt and J. W. F. Valle.

\bibitem{atmo98} 
R. Foot, R.~R. Volkas, O. Yasuda, TMUP-HEL-9801; O.~Yasuda,
Phys. Rev. {\bf D58}, 091301 (1998); G.~L. Fogli, E. Lisi, A. Marrone,
G. Scioscia, Phys. Rev. {\bf D59}, 033001 (1999) hep-ph/9808205;
E.Kh. Akhmedov, A.~Dighe, P. Lipari and A.Yu. Smirnov, hep-ph/9808270

\bibitem{lipari} 
P. Lipari, M. Lusignoli, Phys.\ Rev.\ {\bf D60}, 013003 (1999)
and Phys.\ Rev.\ {\bf D58}, 073005 (1998).

\bibitem{louis}
W. Louis, http://www.neutrino.lanl.gov/LSND/

\bibitem{cobe}
G.~F. Smoot et~al., \apj{396}{92}{L1-L5};
E.L.~Wright et al., \apj{396}{92}{L13}

\bibitem{iras} R. Rowan-Robinson, in {\sl Cosmological Dark Matter},
(World Scientific, 1994), ed. A. Perez, and J. W. F. Valle, p. 7-18,
ISBN 981-02-1879-6

\bibitem{cobe2} 
J.R.~Primack and M.A.~Gross,  astro-ph/9810204; E. Gawiser and J. Silk,
astro-ph/9806197; Science, {\bf 280}, 1405 (1998), and references
therein.

\bibitem{Gelmini:1984pe}
G.~Gelmini, D.N.~Schramm and J.~W.~F.~Valle, Phys. Lett. {\bf 146B},
311 (1984).

\bibitem{ma1} 
J. Bond and G. Efstathiou, \pl{B265}{91}{245}; M. Davis et al.,
\nat{356}{92}{489}; S. Dodelson, G. Gyuk and M. Turner,
\prl{72}{94}{3754}; H. Kikuchi and E. Ma, \pr{D51}{95}{296}; H. B. Kim
and J. E. Kim, \np{B433}{95}{421}; M. White, G. Gelmini and J. Silk,
\pr{D51}{95}{2669};
A.~Masiero, D.~Montanino and M.~Peloso,  hep-ph/9902380.

\bibitem{Hannestad:1999xy}
S.~Hannestad, ``Probing neutrino decays with the cosmic microwave
background," Phys. Rev. {\bf D59}, 125020 (1999) astro-ph/9903475.

\bibitem{veloc} A.G.~Lyne and D.R.~Lorimer, \nat{369}{94}{127}.

\bibitem{Chugai} N.N.~Chugai, {\it Pis'ma Astron. Zh.}{\bf 10}, 87 (1984).

\bibitem{others} A.~Vilenkin, \apj{451}{95}{700}; Dong Lai,
Y.-Z. Qian, astro-ph/9712043

\bibitem{vilenkin98} 
A.~Kusenko, G.~Segre and A.~Vilenkin, ``Neutrino transport: No
asymmetry in equilibrium," Phys. Lett. {\bf B437}, 359 (1998)
astro-ph/9806205.

\bibitem{NSSV} H.~Nunokawa, V.B.~Semikoz, A.Yu.~Smirnov and
J.~W.~F.~Valle, \np{B501}{97}{17}

\bibitem{KusSeg96} A.~Kusenko, G.~Segr\`e, \prl{77}{96}{4872} \&
\prl{79}{97}{2751}; Y.Z.~Qian, \prl{79}{97}{2750}

\bibitem{ALS} 
E.Kh.~Akhmedov, A.~Lanza and D.W.~Sciama, \pr{D56}{97}{6117}

\bibitem{Grasso:1998tt}
D.~Grasso, H.~Nunokawa and J.~W.~F.~Valle, Phys. Rev. Lett. {\bf 81},
2412 (1998), astro-ph/9803002.

\bibitem{Janka:1999kb}
H.T.~Janka and G.G.~Raffelt,  Phys. Rev. {\bf D59}, 023005 (1999)
astro-ph/9808099.

\bibitem{Kusenko:1998bk}
A.~Kusenko and G.~Segre,
Phys.\ Rev.\ {\bf D59} (1999) 061302, astro-ph/9811144.

\bibitem{nuno98} D. Grasso, H. Nunokawa, A . Rossi, A. Yu. Smirnov,
J.~W.~F.~Valle, \ip

\bibitem{fornengo99}
N. Fornengo, M.~C. Gonzalez-Garcia, J.~W.~F.~Valle, 
hep-ph/9906539.

\bibitem{atmdecay} V. Barger et al, hep-ph/9907421 and Phys.\ Rev.\ 
  Lett.\ {\bf 82}, 2640 (1999)
 
\bibitem{exotic}
M. Gasperini, Phys.\ Rev.\ {\bf D38}, 2635 (1988);
J. Pantaleone, A. Halprin, C.N. Leung, Phys.\ Rev.\ {\bf D47}, 4199 (1993); 
A.\ Halprin, C.N. Leung, J. Pantaleone  
Phys.\ Rev.\ {\bf D53}, 5365 (1996).
S. Coleman, S. L. Glashow, Phys.\ Lett.\ {\bf B405}, 249 (1997); 
S. L. Glashow, A. Halprin, P.I. Krastev , C.N. Leung, J. Pantaleone, 
Phys.\ Rev.\ {\bf D56}, 2433 (1997). 
S. Coleman, S. L. Glashow, Phys.\ Rev.\ {\bf D59}, 116008 (1999). 
%
A.M.~Gago, H.~Nunokawa and R.~Zukanovich Funchal,
hep-ph/9909250.

\bibitem{Lola:1998xp}
S.~Lola and J.D.~Vergados, Prog. Part. Nucl. Phys. {\bf 40}, 71
(1998); G.~Altarelli and F.~Feruglio, Phys. Lett. {\bf B439}, 112
(1998) hep-ph/9807353.

\bibitem{Altarelli:1998ns}
G.~Altarelli and F.~Feruglio, Phys. Lett. {\bf B451} (1999) 388
hep-ph/9812475; S.~Lola and G.G.~Ross,  hep-ph/9902283; R.~Barbieri,
L.J.~Hall and A.~Strumia, Phys. Lett. {\bf B445}, 407 (1999),
hep-ph/9808333; M.E.~Gomez, G.K.~Leontaris, S.~Lola and J.D.~Vergados,
Phys. Rev. {\bf D59}, 116009 (1999), hep-ph/9810291; G.K.~Leontaris,
S.~Lola, C.~Scheich and J.D.~Vergados, Phys. Rev. {\bf D53}, 6381
(1996)

\bibitem{Romao:1999up}
J.C.~Romao, M.A.~Diaz, M.~Hirsch, W.~Porod and J.W.~Valle,
hep-ph/9907499.

\bibitem{Quigg:1999ap} C.~Quigg, Introduction to NuFact '99, the
  ICFA/ECFA Workshop on Neutrino Factories Based on Muon Storage
  Rings, hep-ph/9908357.

\bibitem{CP_LBL_FAC} J.~Sato, hep-ph/9910442; O.~Yasuda, hep-ph/9910428;
  H.~Minakata and H.~Nunokawa, Phys.\ Rev.\ {\bf D57} (1998) 4403
  hep-ph/9705208; A.~Romanino,
hep-ph/9909425.

\bibitem{2227}
J. Schechter and  J. W. F. Valle, \pr{D22}{80}{2227}

\bibitem{Pastor:1999iw} S.~Pastor, J.~Segura, V.B.~Semikoz and
  J.W.~Valle, hep-ph/9905405, to be published in Nucl. Phys. B 
  
\bibitem{DEG}
A. Ioannissyan, J.~W.~F. Valle, \pl{B332}{94}{93-99};
B. Bamert, C.P. Burgess, \pl{B329}{94}{289};
D. Caldwell and R. N. Mohapatra, \pr{D50}{94}{3477};
D. G. Lee and R. N. Mohapatra, \pl{B329}{94}{463}; 
A. S. Joshipura, \zp{C64}{94}{31}

\bibitem{Casas:1999ac} J.A.~Casas, J.R.~Espinosa, A.~Ibarra and
  I.~Navarro, hep-ph/9905381; J.~Ellis and S.~Lola, hep-ph/9904279;
  N.~Haba, Y.~Matsui, N.~Okamura and M.~Sugiura,
hep-ph/9908429.

\bibitem{Ma:1999ut}
E.~Ma,
hep-ph/9907503.

\bibitem{nubrane}
R. N. Mohapatra, A. Perez-Lorenzana, hep-ph/9910474;
A. Ioannissyan, J.~W.~F. Valle, in preparation

\bibitem{Giunti:1999hb} C.~Giunti, hep-ph/9909465.

\bibitem{ptvlate} 
Q. Y. Liu, A. Yu. Smirnov,  Nucl.~Phys. {\bf B524} (1998) 505-523;
V. Barger, K. Whisnant and T. Weiler, Phys.Lett.
{\bf B427} (1998) 97-104; S. Gibbons, R. N. Mohapatra, S. Nandi and
A. Raichoudhuri, Phys.~Lett.  {\bf B430} (1998) 296-302;
Nucl.Phys. {\bf B524} (1998) 505-523; S. Bilenky, C. Giunti and
W. Grimus, Eur.  Phys. J. {\bf C 1}, 247 (1998); S. Goswami,
Phys. Rev. {\bf D 55}, 2931 (1997); N. Okada and O. Yasuda,
Int.~J.~Mod.~Phys. {\bf A12} (1997) 3669-3694

\bibitem{smir} E. J. Chun, A. Joshipura and A. Smirnov,
in {\sl Elementary Particle Physics: Present and Future} (World
Scientific, 1996), ISBN 981-02-2554-7; P.~Langacker, Phys. Rev. {\bf
D58}, 093017 (1998); M.  Bando and K. Yoshioka,
Prog. Theor. Phys. {\bf 100}, 1239 (1998)

\bibitem{pvhdm} J. R. Primack, et al.
Phys.~Rev.~Lett. {\bf 74} (1995) 2160

\bibitem{OLDsterilemodel}
J. Schechter and  J. W. F. Valle, \pr{D21}{80}{309}

\bibitem{bbnsterile} 
R. Barbieri and A. Dolgov, Phys. Lett. {\bf B 237}, 440 (1990); 
K. Enquist, K. Kainulainen and J. Maalampi, Phys. Lett. {\bf B 249}, 
531 (1992); D. P. Kirilova and M. Chizov, hep-ph/9707282.

\bibitem{sarkar} S. Sarkar, Rep. Prog. Phys. {\bf 59}, 1493 (1996);
P. J. Kernan and S. Sarkar, \pr{D 54}{96}{R3681}

\bibitem{Fiorentini:1998fv}
G.~Fiorentini, E.~Lisi, S.~Sarkar and F.L.~Villante, ``Quantifying
uncertainties in primordial nucleosynthesis without Monte Carlo
simulations," Phys. Rev. {\bf D58}, 063506 (1998); E.~Lisi, S.~Sarkar
and F.L.~Villante,  Phys. Rev. {\bf D59}, 123520 (1999)

\bibitem{Foot:1997qc}
R. Foot, R.R. Volkas, \pr{D55}{97}{5147-5176};
A.D.~Dolgov, S.H.~Hansen, S.~Pastor and D.V.~Semikoz,
hep-ph/9910444.

\bibitem{SNO} SNO collaboration, E. Norman et al. {\sl Proc. of
The Fermilab Conference: DPF 92} ed.  C. Albright, P. H.  Kasper,
R. Raja and J. Yoh (World Scientific), p.~1450.

\bibitem{vissani} A. Smirnov and F. Vissani, Phys.Lett. {\bf B432}
(1998) 376; J. G.  Learned, S. Pakvasa and J. Stone, hep-ph/9805343;
H.~Murayama and L.~Hall, hep-ph/9806218

\bibitem{JV95}
A. Joshipura, J. W. F. Valle, \np{B440}{95}{647}.

\end{thebibliography}
\end{document}